\documentclass[twocolumn,showpacs,preprintnumbers,amsmath,amssymb,pra,superscriptaddress]{revtex4}
\usepackage{graphicx}
\usepackage{epstopdf}
\usepackage{dcolumn}
\usepackage{bm}
\usepackage{amsmath}
\usepackage{amssymb}
\usepackage{mathrsfs}
\usepackage{amsfonts}
\usepackage{color}

\begin{document}

\title{Fate of photon blockade in the deep strong coupling regime}
\author{Alexandre Le Boit\'e}
\thanks{These authors contributed equally to this work.}
\author{Myung-Joong Hwang}
\thanks{These authors contributed equally to this work.}

\affiliation{Insitut f\"ur Theoretische Physik and IQST, Albert-Einstein-Allee 11, Universit\"at Ulm, D-89069 Ulm, Germany}
\author{Hyunchul Nha}
\affiliation{Department of Physics, Texas A\&M University at Qatar, Education City, P.O.Box 23874,Doha, Qatar}
\affiliation{School of Computational Sciences, Korea Institute for Advanced Study, Seoul 130-722, Korea}
\author{Martin B. Plenio}
\affiliation{Insitut f\"ur Theoretische Physik and IQST, Albert-Einstein-Allee 11, Universit\"at Ulm, D-89069 Ulm, Germany}

\begin{abstract}
We investigate the photon emission properties of the driven-dissipative Rabi model in the so-called ultrastrong and deep strong coupling regimes, where the atom-cavity coupling rate $g$ becomes comparable or larger than the cavity frequency $\omega_c$.  By solving numerically the master equation in the dressed-state basis, we compute the output field correlation functions in the steady-state for a wide range of coupling rates. We find that, as the atom-cavity coupling strength increases, the system undergoes multiple transitions in the photon statistics. In particular, a first sharp anti-bunching to bunching transition, occurring at $g\sim0.45\omega_c$, leading to the breakdown of the photon blockade due to the counter-rotating terms, is shown to be the consequence of a parity shift in the energy spectrum. A subsequent revival of the photon blockade and the emergence of the quasi-coherent statistics, for even larger coupling rates, are attributed to an interplay between the nonlinearity in the energy spectrum and the transition rates between the dressed states.
\end{abstract}
\pacs{42.50.Ar, 03.67.Lx, 42.50.Pq, 85.25.-j}

\maketitle
\section{Introduction}
Cavity quantum electrodynamics (cavity QED) has proved to be a powerful platform both for testing fundamental laws of quantum physics \cite{Haroche:2006, Gleyzes:2007} and implementing quantum information protocols \cite{Imamoglu:1999,Majer:2007}.  A strong atom-cavity interaction, larger than any dissipation rates, can be readily achieved in cavity QED setups \cite{Rempe:1987, Reithmaier:2004, Wallraff:2004}, giving rise to interesting nonlinear effects induced by effective photon-photon interactions. One of the most famous is the photon blockade effect~\cite{Imamoglu:1997}, where the presence of a single photon inside the cavity is sufficient to inhibit the absorption of another photon. The photon blockade results in a highly nonclassical emission of photons from the cavity, characterized by a strong antibunching. The photon blockade has been investigated theoretically and experimentally in a large variety of systems, such as atomic cavity QED~\cite{Birnbaum:2005,Carmichael:2015}, semiconductor nanostructures~\cite{Verger:2006, Liew:2010} and superconducting circuits~\cite{Bozyigit:2010, Lang:2011,Hoffman:2011}.

The effective photon-photon interaction in a cavity QED system is well captured, in the strong coupling regime, by the Jaynes-Cummings (JC) model. Its energy spectrum has a nonlinear ladder structure consisting of a doublet of energy eigenstates with the same number of excitation $n$. This nonlinearity is responsible for the photon blockade. More precisely, the avoided crossing between two states forming a doublet shows a characteristic scaling of $g\sqrt{n}$, where $g$ is the atom-cavity coupling strength \cite{Fink:2008}. Therefore, the JC model predicts that, for a fixed number of excitations, the nonlinearity of the energy spectrum increases with the coupling strength and the photon-blockade effect gets more pronounced.

Meanwhile, recent experimental progress in tailoring the light-matter interaction has made it possible to achieve a coupling strength that is comparable or even larger than the cavity frequency~\cite{Devoret:2007, Bourassa:2009, Todorov:2010, Niemczyk:2010, Forn-Diaz:2010, Nataf:2011, Forn-Diaz:2016, Yoshihara:2016}. In this so-called ultrastrong coupling regime, the rotating-wave approximation on which the Jaynes-Cummings model is based is no longer valid and thus the counter-rotating terms cannot be neglected. A description of the cavity QED system based on the Rabi model, whose spectrum and eigenstates have been a subject of intensive studies \cite{Irish:2007, Ashhab:2010, Hwang:2010, Braak:2011, Nataf:2011,Hwang:2015}, is therefore necessary.  In order to investigate output photon correlations, which are experimentally relevant for many setups, several works have also considered driven-dissipative scenarios \cite{Ciuti:2006,DeLiberato:2009,Beaudoin:2011,Ridolfo:2012, Henriet:2014}. They have shown that in the ultrastrong coupling regime, the usual quantum optical master equation and input-output relations led to unphysical predictions and had to be modified.

In this context, the question of the fate of the photon-blockade effect in the ultrastrong coupling regime naturally arises. It was shown in Ref.~\cite{Ridolfo:2012} that there is no qualitative change in the output-photon statistics when the coupling strength is of the order of $10\%$ of the cavity frequency, except when a bias on the qubit, which breaks explicitly the parity symmetry of the Rabi model, is introduced. In the latter case, an output-photon statistics that is significantly different from the standard scenario based on the JC model is predicted. The effect of counter-rotating terms on the output-photon statistics and the photon blockade when the coupling strength is of the same order of the cavity frequency~\cite{Larson:2007, Casanova:2010, Hwang:2016, Yoshihara:2016}, however, has remained largely unexplored.

In this paper, we study the photon emission properties of a cavity QED system in the ultrastrong coupling regime for a wide range of atom-cavity coupling strength $g$, up to several multiples of the cavity frequency $\omega_c$. To this end, we investigate the steady-state properties of the Rabi model in a driven-dissipative setting. We consider a weak excitation limit where the driving field intensity is smaller than the dissipation rates of the cavity and the atom. By solving the  master equation in the dressed state basis, we compute the output field correlation functions in the steady-state and show that, as the coupling strength increases, the counter-rotating terms lead to a breakdown and revival of the photon blockade, followed by the emergence of a quasi-coherent photon statistics. We explain the multiple transitions in the photon statistics due to the counter-rotating terms, by analyzing the energy spectrum of the Rabi Hamiltonian and the properties of the dressed states. In particular, the first breakdown of the photon blockade, evidenced by a sharp antibunching to bunching transition in the photon statistics, occurs at $g\sim0.45\omega_c$ and it is attributed to the parity shift occurring in the energy spectrum, which induces a cascaded emission of photons.

The paper is organized as follows: the theoretical framework is presented in Sec.~\ref{sec:model}. Section~\ref{sec:res} is devoted to the main numerical results and key observations. More detailed discussions on the three different phases of photon emission are divided into three subsections: the break-down and revival of the photon-blockade effect (Secs. \ref{subsec:break} and \ref{subsec:revival} respectively) and the transition to a coherent regime (\ref{subsec:coherent}). We conclude in Sec. \ref{sec:conclu}.

%
%

\section{Theoretical Framework}
\label{sec:model}
We consider the Rabi Hamiltonian, describing a single cavity mode coupled to a two-level atom,
\begin{equation}
\label{Hamilto_r}
H_r = \omega_c a^{\dagger}a + \omega_a\sigma_+\sigma_- -g(a+a^{\dagger})\sigma_x,
\end{equation}
where we have introduced the photon annihilation operator $a$, and the Pauli matrices $\sigma_x$, $\sigma_y$ (with $\sigma_{\pm}  = \frac{1}{2}(\sigma_x \pm i\sigma_y)$). Here, $\omega_c$ is the cavity frequency, $\omega_a$ the atomic transition frequency, and $g$ the atom-cavity coupling strength. In the following we will focus on a resonant case, i.e., $\omega_c = \omega_a$.  An important property of the Rabi Hamiltonian is that the parity of the total number of excitations, $\Pi=\exp[i\pi(a^\dagger a +\sigma_+\sigma_-)]$, is a conserved quantity. It is therefore convenient to label the eigenstates of the system using the notation $|\Psi_j ^{\pm}\rangle$ for the $j^{th}$ eigenstate ($j=0,1,..$) of the $\pm$ parity subspace and $E_{j}^{\pm}$ for the corresponding energy. For example, the ground state of $H_r$ is $|\Psi_0^{+}\rangle$, which is the lowest energy state of the $+$ parity subspace; the first excited state of $H_r$, which corresponds to the lowest energy state of the $-$ parity subspace, is $|\Psi_0^{-}\rangle$.

We focus in this paper on a driven-dissipative scenario where the cavity is externally driven by a monochromatic field and both the cavity and the atom are coupled to their environments, leading to dissipation. The total time-dependent Hamiltonian of the system is
\begin{equation}\label{Hamilto}
H(t) = H_r + F\cos(\omega_dt)(a+a^{\dagger}),
\end{equation}
with $F$ the intensity of the driving field and $\omega_d$ its frequency.
The dynamics of this open quantum system is governed by a master equation of the form,
\begin{equation}
\label{ME}
\partial_t \rho = i[\rho,H(t)] +\mathcal{L}_a\rho+\mathcal{L}_{\sigma}\rho,
\end{equation}
where the term $\mathcal{L}_a\rho+\mathcal{L}_{\sigma}\rho$ describes the dissipation of the system excitations into the electromagnetic environment. In the standard quantum optical master equation, the quantum jump operators appearing in the term $\mathcal{L}_a\rho+\mathcal{L}_{\sigma}\rho$ are given by the annihilation operator of the cavity $a$ and the atom $\sigma_-$; the validity of this approach relies on the fact that the atom and the cavity can be regarded as being independently coupled to their own environment. In the ultrastrong coupling regime, this assumption is no longer valid and the dissipation should be described as quantum jumps among the eigenstates of the total atom-cavity Hamiltonian~\cite{Beaudoin:2011}. A correct master equation, taking fully into account the coherent coupling between the cavity and the atom, can therefore be expressed in the dressed-state basis $\{|\Psi_j^{p}\rangle \}$ with $p=\pm$, in which the Hamiltonian (without driving) is diagonal. In this basis, the dissipative part reads,
\begin{align}
\mathcal{L}_a\rho+\mathcal{L}_{\sigma}\rho=\sum_{p=\pm}\sum_{k,j}\Theta(\Delta_{jk}^{p\bar{p}})\left(\Gamma_{jk}^{p\bar{p}}+K_{jk}^{p\bar{p}}\right)\mathcal{D}[|\Psi_j^{p}\rangle\langle \Psi_k^{\bar{p}}|],
\end{align}
where $\Theta(x)$ is a step function, i.e., $\Theta(x)=0$ for $x\leq0$ and $\Theta(x)=1$ for $x>1$, and $\bar{p}=-p$.
We have also introduced the following notation, $\mathcal{D}[\mathcal{O}] = \mathcal{O}\rho\mathcal{O}^{\dagger} - \frac{1}{2}(\rho\mathcal{O}^{\dagger}\mathcal{O} + \mathcal{O}^{\dagger}\mathcal{O}\rho)$.
The quantities $\Gamma_{jk}^{p\bar{p}}$ and $K_{jk}^{p\bar{p}}$ denote the rates of transition from a dressed-state $|\Psi_k^{\bar{p}}\rangle$ to $|\Psi_j^{p}\rangle$ due to the atomic and cavity decay, respectively; the transition rates are defined as

\begin{align}\label{trans_rates}
\Gamma_{jk}^{p\bar p} &= \gamma\frac{\Delta_{jk}^{p \bar p}}{\omega_c}|\langle \Psi_j^{p}|(a - a^{\dagger})|\Psi_k^{\bar p}\rangle|^2, \\
 K_{jk}^{p\bar p} &= \kappa\frac{\Delta_{jk}^{p\bar p}}{\omega_c}|\langle \Psi_j^{p}|(\sigma_- - \sigma_+)|\Psi_k^{\bar p}\rangle|^2,
\end{align}
where $\Delta_{jk}^{p\bar p} = E_k^{\bar p} - E_j^{p}$ is the transition frequency and $\gamma$, $\kappa$ are respectively the cavity and the atom decay rates. Note that the transition between states belonging to the same parity space is forbidden because both operators $a - a^{\dagger}$ and $\sigma^- - \sigma^+$ change the parity of the state. 

In the following, we will be interested in the long time dynamics of Eq.~(\ref{ME}). Except for a sufficiently small $g$, Eq~(\ref{ME}) generally does not have a particular rotating-frame where the equation becomes time-independent. Therefore, the solution has a residual oscillation at the driving frequency $\omega_d$ even in the $t\rightarrow\infty$ limit. The steady-state properties are then obtained by averaging the solution over several driving periods, which corresponds to a time integrated measurement in an actual experiment~\cite{Ridolfo:2012}.

%

\begin{figure}
 	\centering
	\includegraphics[width = \columnwidth]{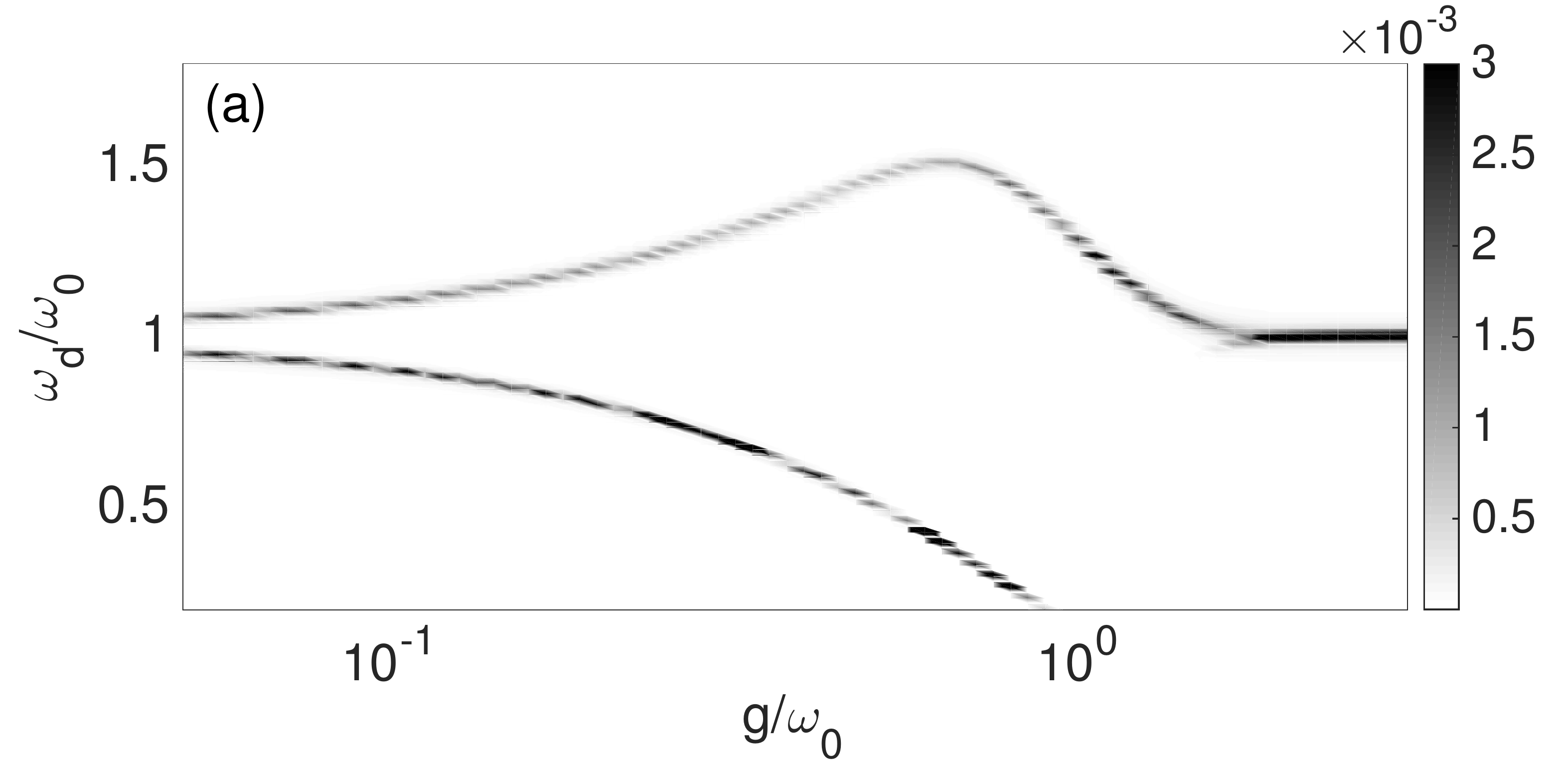}\\
	\includegraphics[width = \columnwidth]{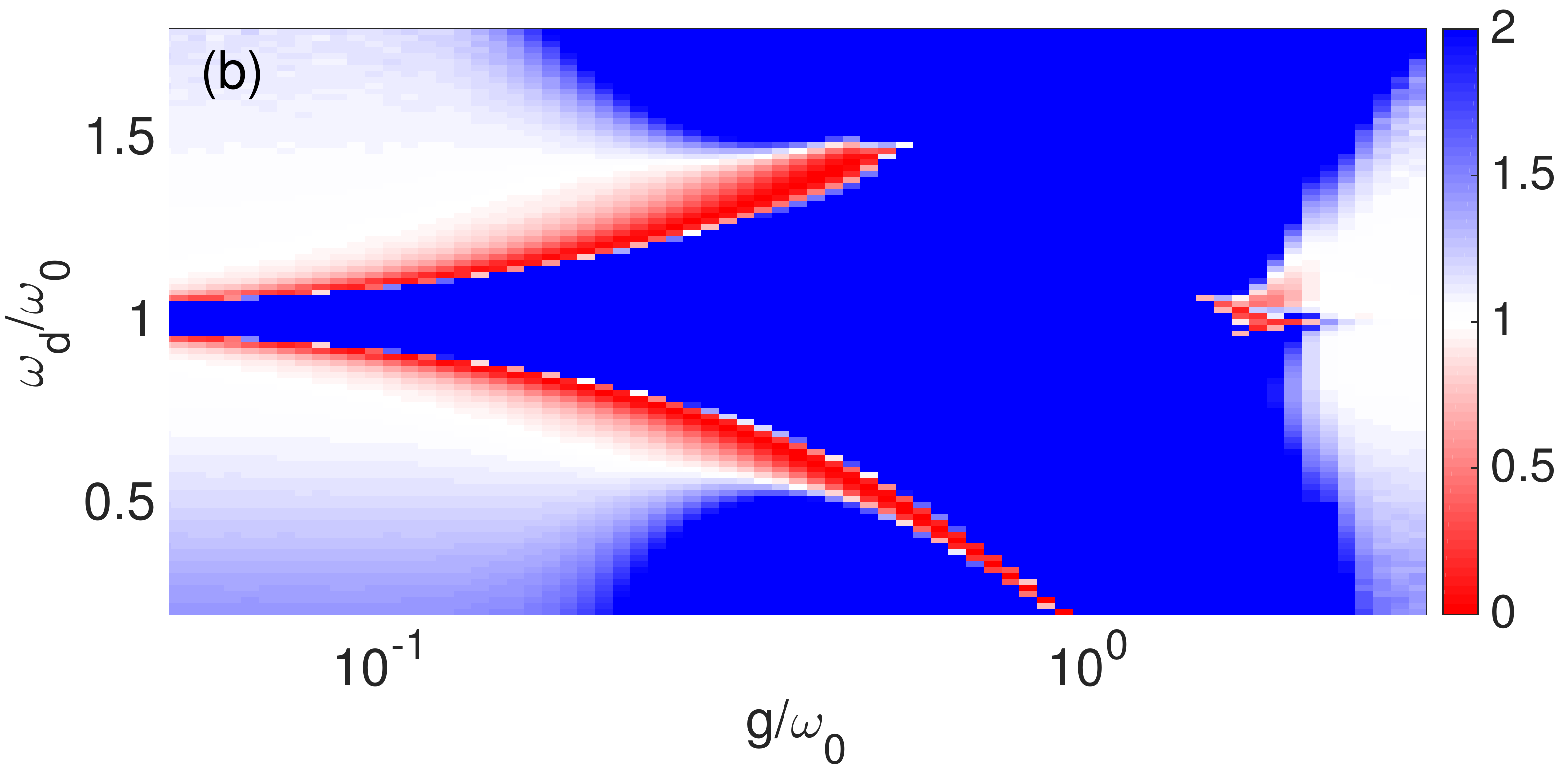}
\caption{(color online). (a) Intensity $I_{out}$  as a function of $g/\omega_c$,  and $\omega_d/\omega_c$. (b) Second order correlation function $g^{(2)}(0)$. To improve readability, the color scale for the autocorrelation function saturates at $g^{(2)}(0) = 2$. The dissipation and driving parameters are $\gamma/\omega_c = 10^{-2}$ and $F/\omega_c = 10^{-3}$.}
\label{fig:diag}
\end{figure}

\section{Results}
\label{sec:res}
Our aim is to study the statistics of the photons emitted from the cavity as a function of the coupling strength $g$ in the weak excitation limit ($F/\gamma, F/\kappa \ll 1$). When the atom-cavity coupling strength is sufficiently small, the output field is proportional to the intracavity field $a$, and the correlation functions of the intracavity field are therefore the relevant observables for characterizing the emitted light. This is not the case in the ultrastrong coupling regime and the standard input-output relation is modified~\cite{Ciuti:2006,DeLiberato:2009,Ridolfo:2012}. In this case, as shown in Ref.~\cite{Ridolfo:2012}, the output field is proportional to an operator $\dot X^+$, defined in the dressed-state basis as:
\begin{equation}
\dot{X}^+  = \sum_{p=\pm}\sum_{k,j}\Theta(\Delta_{jk}^{p\bar{p}}) \Delta_{jk}^{p\bar{p}}|\Psi_j^{p}\rangle \langle \Psi_j^{p}|i(a^\dagger-a)|\Psi_k^{\bar{p}}\rangle \langle \Psi_k^{\bar{p}}|.
\end{equation}
The second-order correlation function then reads
\begin{equation}
g^{(2)}(0) = \frac{\langle \dot X^-\dot X^-\dot X^+ \dot X^+\rangle}{\langle \dot X^-\dot X^+\rangle^2},
\end{equation}
and the intensity of the emitted photons is proportional to $I_{out}=\langle \dot X^-\dot X^+\rangle$. A strong photon antibunching indicated by $g^{(2)}(0)\ll 1$ will be the signature of the photon-blockade effect.

\begin{figure}
 	\centering
	\includegraphics[width = 0.49\columnwidth]{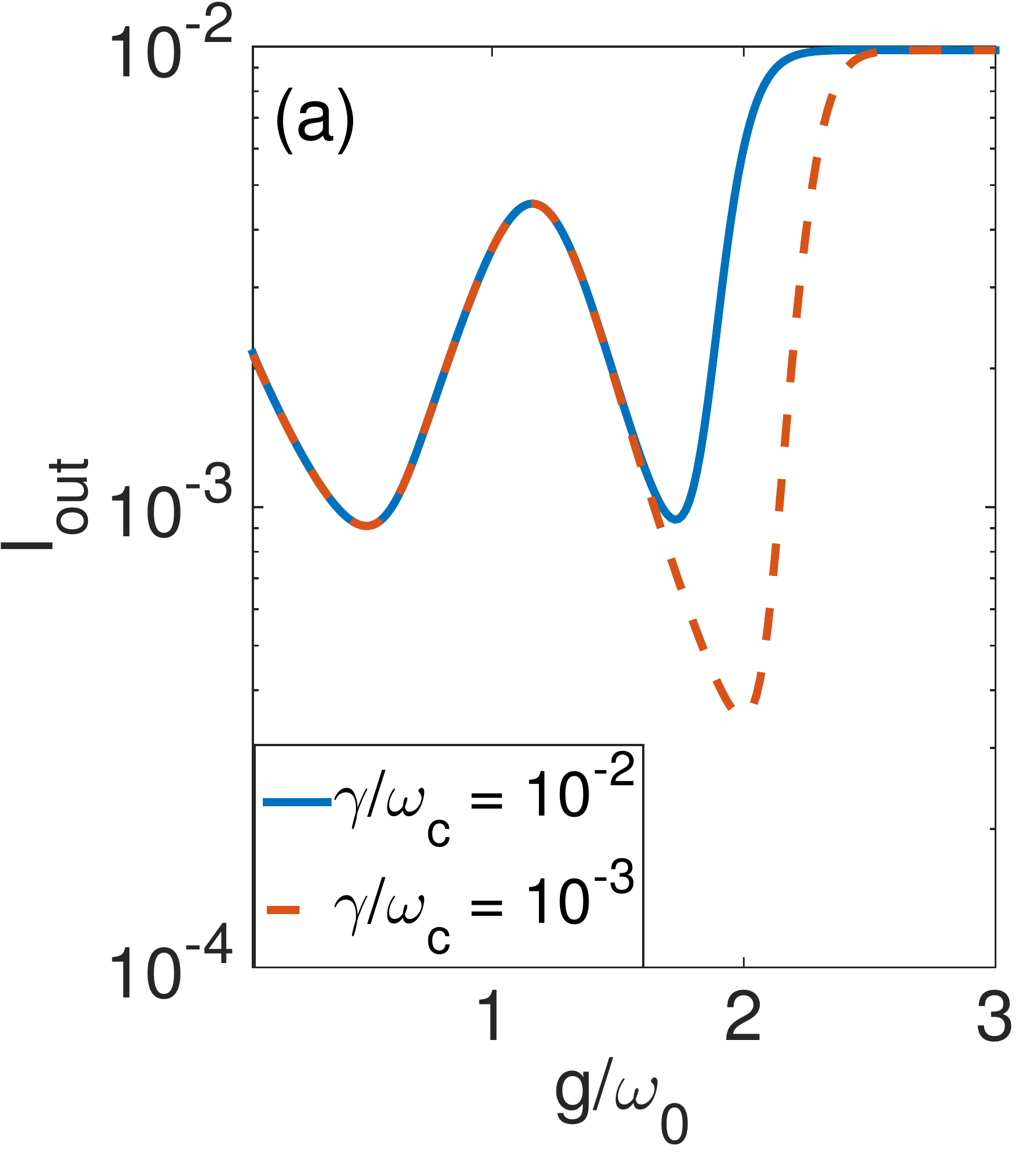}
	\includegraphics[width = 0.49\columnwidth]{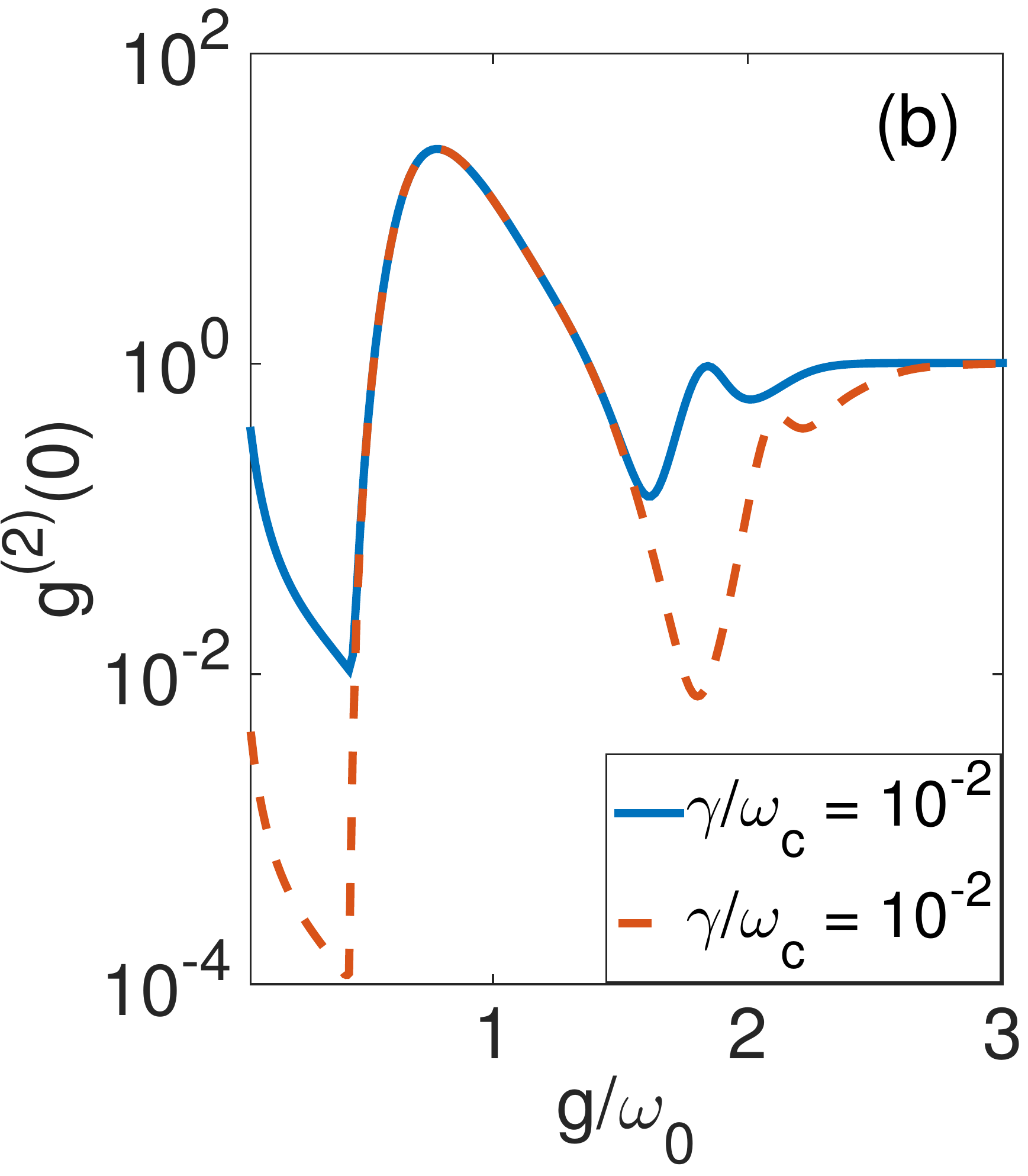}
\caption{(color online). (a) Intensity $I_{out}$ and (b) Second order correlation function $g^{(2)}(0)$  as a function of $g/\omega_c$, for an external driving resonant with the transition $|\Psi_0^+\rangle \to |\Psi_1^-\rangle$. Results are shown for two values of the dissipation rate, $\gamma/\omega_c = 10^{-2}$ (solid blue line) and $\gamma/\omega_c = 10^{-3}$ (dashed red lines). In both cases $F/\gamma = 10^{-1}$.}
\label{fig:gscan}
\end{figure}

Figure~\ref{fig:diag} shows a phase diagram for $I_{out}$ and $g^{(2)}(0)$ in the steady-state as a function of the coupling strength and the driving frequency. They are obtained by solving Eq.~(\ref{ME}) numerically for a fixed driving strength and dissipation rates, $F/\gamma = 10^{-1}$ and $\gamma/\omega_c = \kappa/\omega_c =  10^{-2} $. The two dark lines in Fig.~\ref{fig:diag} (a) correspond to a resonant driving of the two transitions $|\Psi_0^+\rangle\to|\Psi_0^-\rangle $ and $|\Psi_0^+\rangle\to|\Psi_1^-\rangle $.
According to the standard photon blockade scenario, a strong anti-bunching occurs for both driving frequencies. 
This persists even when one enters the ultrastrong coupling regime ($g/\omega_c \approx 10^{-1}$), where it is well-known that the rotating-wave approximation breaks down~\cite{Ridolfo:2012}. In this range of the coupling strength, the counter-rotating terms make the photon-blockade effect more pronounced as evidenced by the decreasing $g^{(2)}(0)$.

Interestingly, however, as the coupling strength increases further, entering the so-called deep strong coupling regime ($g/\omega_c \approx 1$), the photon statistics of the first transition and the second transition become drastically different. Regarding the first transition $|\Psi_0^+\rangle\to|\Psi_0^-\rangle$, the emitted light remains strongly antibunched. This feature can be easily understood from the fact that, in the deep strong coupling regime, the lowest two eigenstates of the Rabi model $|\Psi_0^+\rangle$ and $|\Psi_0^-\rangle$ becomes progressively closer in energy as $g$ increases, eventually forming an effective two-level system that is separated from the higher energy states by the cavity frequency $\omega_c$(for $g/\omega_c \gtrsim1$)~\cite{Nataf:2011}. The nonlinearity for the first transition therefore increases monotonically as a function of $g$, and hence the photon blockade effect becomes more pronounced. It is important to note however that the intensity $I_{out}$ of the emitted photon goes to zero for the first transition for $g/\omega_c\gtrsim 1$, which eventually sets a practical limit for the observation of the photon blockade effect by driving this transition.

Richer and more interesting properties of the emitted photons are observed when driving the second transition. The most prominent feature is that the dip in $g^{(2)}(0)$ disappears for $0.45\lesssim g/\omega_c\lesssim1$. The dip briefly reappears for a larger value of $g$ before the emitted light eventually becomes coherent $g^{(2)}(0)=1$. For a more precise understanding, we present in Fig.~\ref{fig:gscan} a cut of the phase diagrams of Fig.~\ref{fig:diag} following the second transition.

Overall, in terms of statistics of the emitted photons, there appears three distinctive phases in the deep strong coupling regime: (a) For $0.45\lesssim g \lesssim1$, there occurs a sharp transition from antibuncing to bunching leading to the breakdown of the photon blockade (b)  For $1\lesssim g \lesssim2.5$, another bunching-to-antibunching transition leads to the revival of the photon blockade. (c) For $ g \gtrsim 2.5$, $g^{(2)}(0)$ converges to 1, that is the emitted light becomes coherent. This indicates that the statistics of emitted photons reverts back to that of a linear cavity for coupling strengths that are multiple times larger than the cavity frequency. The mechanisms giving rise to these three different kinds of photon statistics are analyzed in more details in the remainder of this section.

\subsection{Breakdown of the photon blockade ($0.45\lesssim g/\omega_c\lesssim1$)}
\label{subsec:break}
\begin{figure}
 	\includegraphics[width = 0.98\columnwidth]{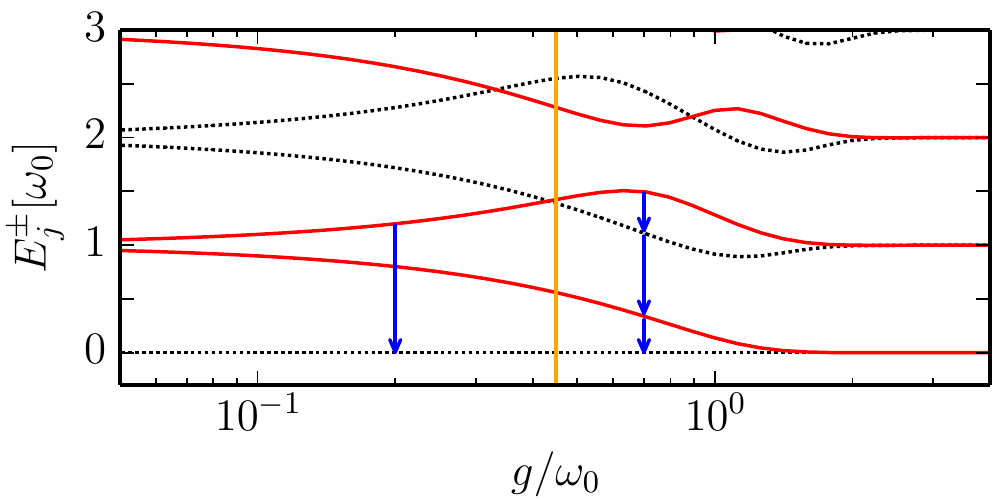}\\
	\includegraphics[width = 0.49\columnwidth]{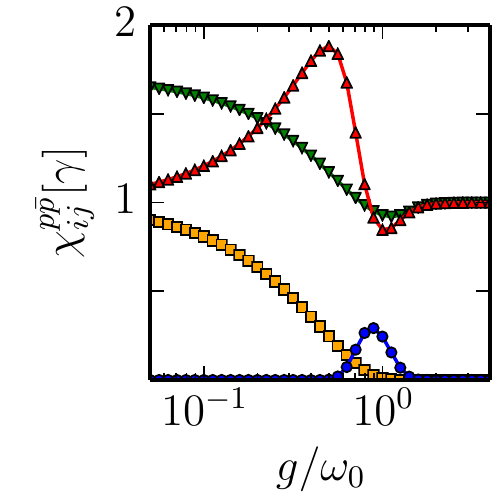}
	\includegraphics[width = 0.49\columnwidth]{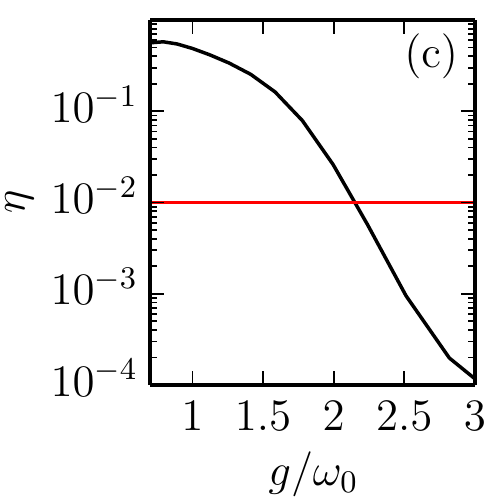}
\caption{(color online) (a) Energy spectrum of the Rabi Hamiltonian (without driving). Black dotted lines indicate energy levels with an even number of excitations while red solid lines correspond to an odd number of excitations. The parity shift of the second excited state is denoted by an horizontal line. Blue arrows show the available decay channels when driving the second transition $|\Psi_{0}^+\rangle\to|\Psi_{1}^-\rangle$. (b) Transition rates between the different dressed states, $\chi^{+-}_{00}$ (yellow squares), $\chi^{+-}_{11}$ (blue circles), $\chi^{-+}_{01}$ (inverted green triangles) and $\chi^{+-}_{01}$ (red triangles), as a function of the coupling strength. 
(c) Anharmonicity of the Rabi Hamiltonian, measured by the quantity $\Delta_{21}^{+-}-\Delta_{10}^{-+}$. The intersection with the vertical line shows the value of $g$ for which the anharmonicity is equal to $\gamma$ (here, $\gamma/\omega_c = 10^{-2}$).}
\label{fig:spec}
\end{figure}

Generally speaking, two main conditions need to be fulfilled in order to observe a photon antibunching when driving the second transition $|\Psi_{0}^+\rangle\to|\Psi_{1}^-\rangle$: i) anharmonicity of the spectrum, ii) no cascaded emission of photons. In order to understand the sharp transition to the photon bunching in this range of coupling strength, we investigate the structure and the selection rules of the energy spectrum of the Rabi model, which is presented on Fig. \ref{fig:spec} (a). One key element is the change in parity for the second excited state, indicated by a vertical line. It occurs for a certain coupling strength $g_c \approx 0.45 \omega_c$ that coincides with the value where the sharp transition occurs in Fig.~\ref{fig:gscan} (b). For $g<g_c$, $|\Psi_1^-\rangle$ is the second excited state, which means that both the first and second excited states belong to the - parity subspace. In this case, the only decay channel available is the direct transition $|\Psi_1^-\rangle \to |\Psi_0^+\rangle$. For $g>g_c$, however, $|\Psi_1^-\rangle$ is the third excited state and the second excited state now belongs to the $+$ parity subspace. An important consequence is that, above $g_c$, two decay channels are available: the first one is simply the direct transition $|\Psi_1^-\rangle \to |\Psi_0^+\rangle$ that results in the emission of a single photon, and the second one is $|\Psi_1^-\rangle \to |\Psi_1^+\rangle \to |\Psi_0^-\rangle \to |\Psi_0^+\rangle$.

More insight on the photon statistics can be gained by investigating the transition rates between the different dressed-states $\chi^{-+}_{ij} =\Gamma^{-+}_{ij} + K^{-+}_{ij} $.
 Note that these transition rates depend on the energy eigenvalues and eigenstates through Eq.~(\ref{trans_rates}) and they are propotional to the dissipation rates $\gamma$  and $\kappa$. In Fig.~\ref{fig:spec} (a), the rates for all the transitions involved when pumping the second transition are plotted. Since the transition rate $\chi^{-+}_{10}$ is much larger than $\chi^{-+}_{11}$, the direct transition channel, $|\Psi_1^-\rangle \to |\Psi_0^+\rangle$, is the dominant one. But whenever the system relaxes through the second channel, $|\Psi_1^-\rangle \to |\Psi_1^+\rangle \to |\Psi_0^-\rangle \to |\Psi_0^+\rangle$, it triggers a cascaded photon emission. Hence, this second channel introduces fluctuations in the photon statistics and is responsible for the strong photon bunching. The value of $g$ for which $g^{(2)}(0)$ is the highest (see Fig.~\ref{fig:gscan} (b)) corresponds indeed to the value of $g$ where the transition rate $\chi^{-+}_{11}$ is the highest.
\begin{figure}[h]
 	\centering
	\includegraphics[width = 0.49\columnwidth]{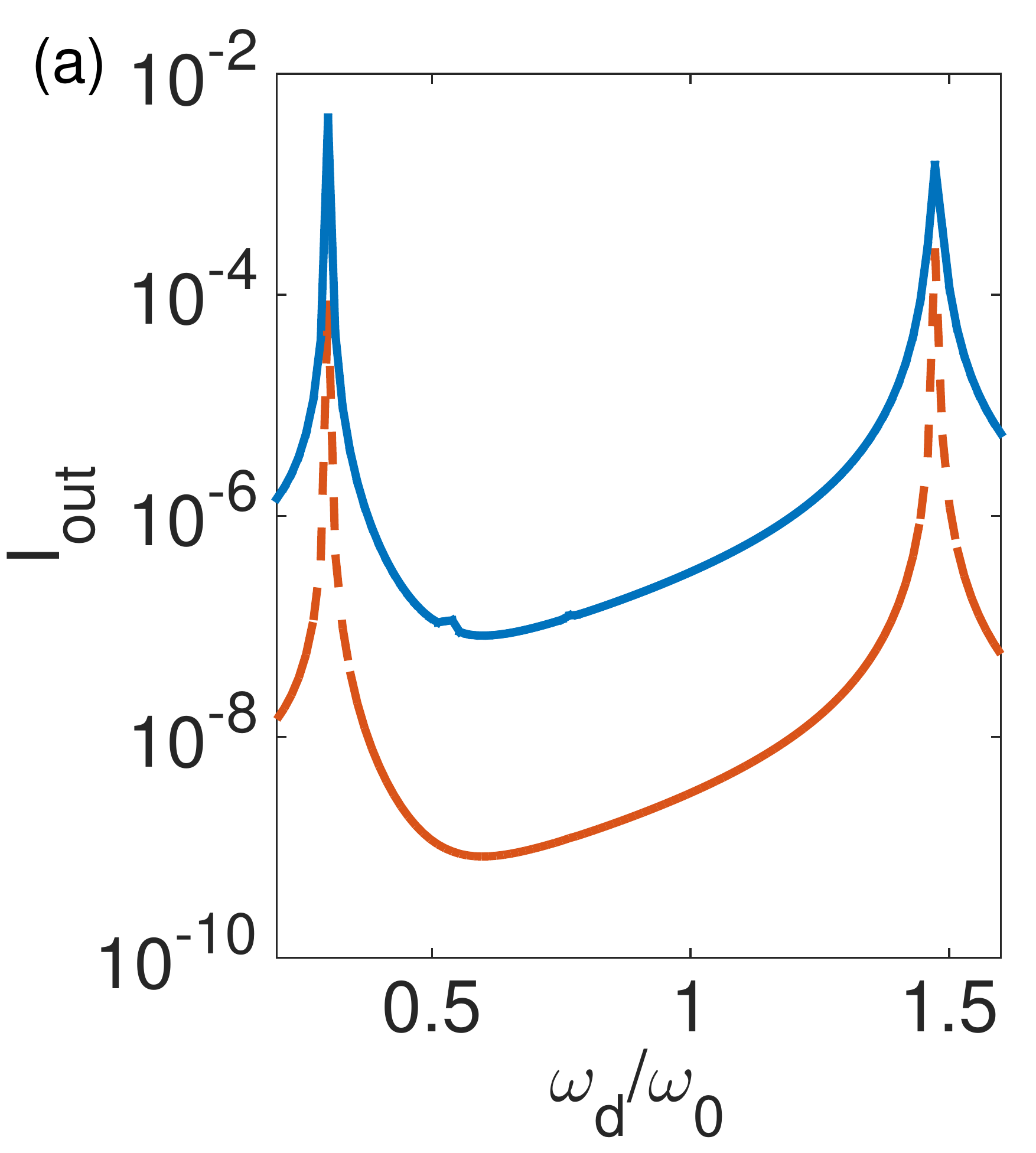}
	\includegraphics[width = 0.49\columnwidth]{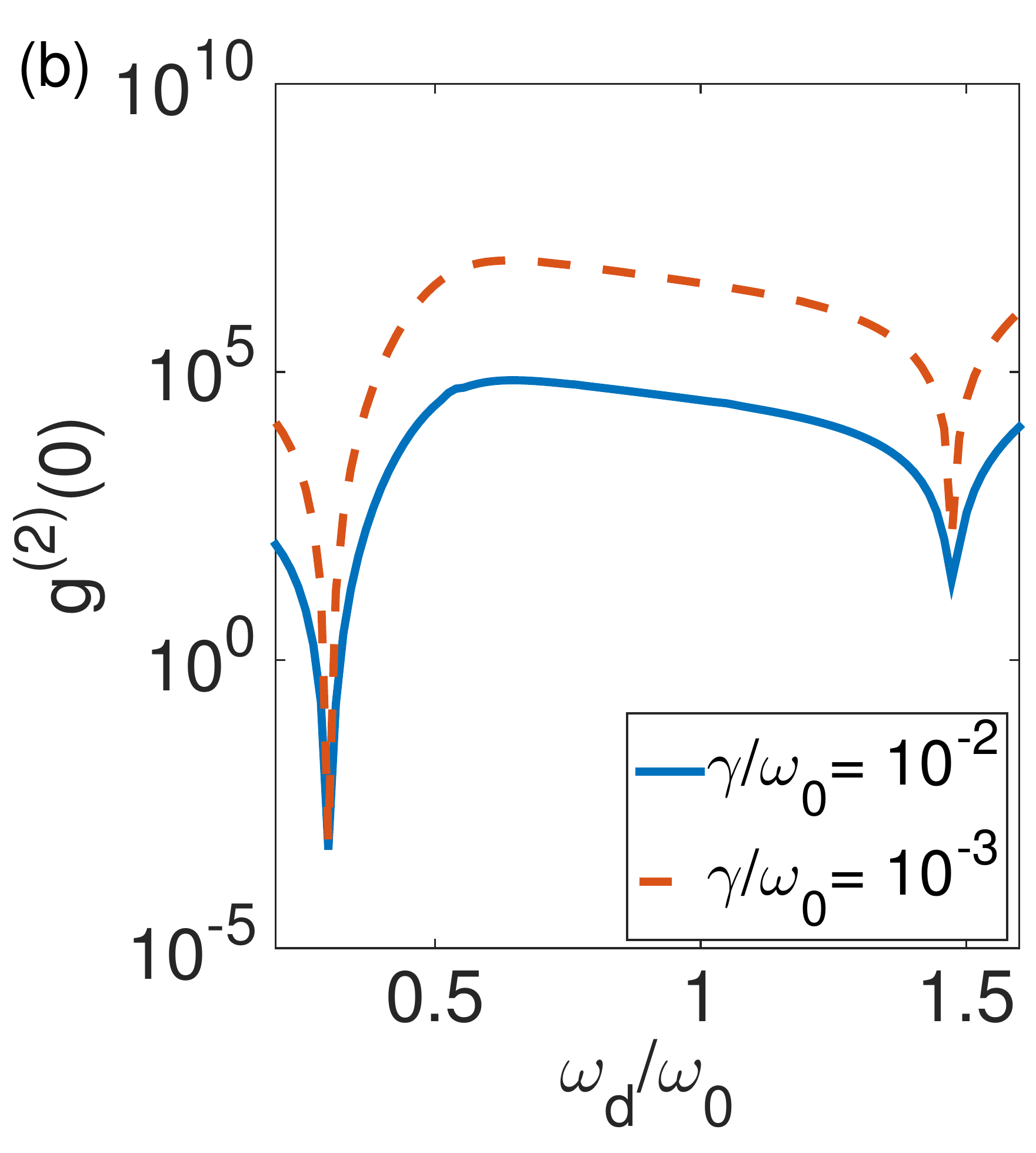}
\caption{(color online). (a) Intensity $I_{out}$  and (b) $g^{(2)}(0)$ as a function of the driving frequency $\omega_d$, for $g/\omega_c = 0.75$. $F$ and $\gamma$ are the same as in Fig.(\ref{fig:gscan}).}
\label{fig:freqscan_g075}
\end{figure}

For a fixed value of $g$, the effect of the change in the level structure can be also seen by looking at $I_{out}$ and $g^{(2)}(0)$ as a function of the pump frequency $\omega_d$. An example of such a plot for $g/\omega_c = 0.75$ is shown in Fig.~\ref{fig:freqscan_g075}. The other parameters are the same as in Fig.~\ref{fig:gscan}. The two dips in $g^{(2)}(0)$ correspond to the two transitions $|\Psi_0^+\rangle \to |\Psi_0^-\rangle$ and $|\Psi_0^+\rangle \to |\Psi_1^-\rangle$, and they match the two peaks in the photon intensity. In terms of photon statistics, the contrast between the first and the second transition is particularly striking. Driving resonantly the first transition results in near-perfect antibunching, while a resonant excitation of the second transition leads to a very strong bunching ($g^{(2)}(0)\gtrsim 10$). Note that the photon bunching is even more pronounced when the external laser is off-resonant. The intensity of the photon is however extremely low in this case. A similar ``superbunching'' behavior for weak and non-resonant excitation has also been predicted in the Jaynes-Cummings Hamiltonian \cite{Shamailov:2010} and for the Kerr Hamiltonian \cite{LeBoite:2014}.

\subsection{Revival of the photon blockade ($1\lesssim g/\omega_c\lesssim2.5$)}
\label{subsec:revival}

Fig.~\ref{fig:spec} (b) shows that for $g/\omega_c \gtrsim 1$, the transition rate $\chi^{+-}_{11}$ decreases and goes eventually to zero.  In other words, the first decay process of the cascaded decay channel closes. As a result, the direct transition becomes again the only available decay channel, and the fluctuations in the photon statistics start to diminish. The closing of the second decay channel is related to the fact that the transition amplitudes $\chi^{p\bar p}_{ij}$ are proportional to the energy spacing $\Delta_{ij}^{p\bar p}$. Indeed, for $g>1$, all the transition frequencies $\Delta^{p\bar p}_{ii}$, and in particular $\Delta^{-+}_{11}$, decreases exponentially with $g/\omega_c$~\cite{Hwang:2016}. This is clearly visible on the energy spectrum presented in Fig. \ref{fig:spec} (a); for large values of $g/\omega_c$, the energy spectrum can be seen as a collection of doublets of quasi-degenerate eigenstates $\{|\Psi^-_{j}\rangle, |\Psi^+_{j}\rangle\}$. 
 
 At the same time, the spacing between the doublets of quasi-degenerate states becomes constant (equal to the cavity frequency $\omega_c$) when $g$ increases, suppressing the anharmonicity of the spectrum. The resulting photon statistics in this regime thus depends on an interplay between the closing of the second decay channel and the vanishing anharmonicity. More specifically, a revival of the photon blockade effect can occur if the second decay channel is closed while a sufficiently large anharmonicity still remains in the energy spectrum. Fig.~\ref{fig:spec} (c), where the anharmonicity $\eta = \Delta_{21}^{+-}-\Delta_{10}^{-+}$ is compared to the dissipation rate $\gamma$ as a function of $g$, shows that this is indeed the case for $1\lesssim g/\omega_c\lesssim2.5$, providing a clear explanation for the revival of the photon blockade. Note that this mechanism also explain why the region for which photon antibunching is recovered becomes larger for smaller values of $\gamma$; see the red dotted line in Fig.~\ref{fig:gscan} (b).  
\subsection{Reversion to non-interacting photons  ($ g/\omega_c \gtrsim 2.5$)}
\label{subsec:coherent}
%
The photon correlation in this regime converges to $g^{(2)}(0)=1$. It is interesting to observe that a very strong atom-cavity coupling leads to a poissonian statistics, which is characteristic of photons emitted from a linear cavity. To understand this rather counter-intuitive observation, it is illuminating to consider the limiting case of  $g/\omega_c \to \infty$~ \cite{Hwang:2016}. In this limit, the energy spectrum is that of an displaced harmonic oscillator and therefore it is harmonic with a characteristic frequency of $\omega_c$. The doubly degenerate $j$-th eigenstates are given by $|j,\pm g/\omega_c\rangle|\pm\rangle$ where $|j, \alpha\rangle = e^{\alpha a^{\dagger}-\alpha^*a}|j\rangle$, with $\alpha$ a complex number, is the $j$-th displaced Fock state, and $\sigma_x|\pm\rangle=\pm|\pm\rangle$. For large but finite values of $g/\omega_c$, the term $\omega_a\sigma_+\sigma_-$ lifts the degeneracy between $|j,\pm g/\omega_c\rangle|\pm\rangle$, leading to an exponentially supressed energy splitting, $\langle n,-g/\omega_c|n,g/\omega_c\rangle \propto e^{-2g^2/\omega_c^2}L_n(4g^2/\omega_c^2)$, where $L_n$ is the $n^{\mathrm{th}}$ Laguerre polynomial.

If the exponentially suppressed anharmonicity of the spectrum for $g/\omega_c\gg1$ is smaller than the dissipation rate $\gamma$, the system will essentially behave as a linear system, and the emitted light will show a Poissonian statistics. Fig.~\ref{fig:spec} (c) clearly shows that the anharmonicity $\eta = \Delta_{21}^{+-}-\Delta_{10}^{-+}$ becomes comparable to the dissipation rate $\gamma$ at $g\sim2$, and it becomes smaller for larger values of $g$. This is consistent with the range of coupling strength where the reversion to non-interacting photons occurs, i.e., $g/\omega_c \gtrsim 2.5$. We emphasize that the mechanism behind the breakdown of the photon blockade in this regime differs fundamentally from the one reported in section \ref{subsec:break}. In the latter case it occurs due to a parity shift in the spectrum and $g_c$ does not depend on $\gamma$ or any other external parameter (see Fig.~\ref{fig:gscan} (b)).

\section{Conclusion}
\label{sec:conclu}
In this paper, we have investigated the output photon statistics of a cavity coupled to a two-level atom in the ultrastrong and deep strong coupling regimes. We have shown that, when the second available transition is driven in the weak excitation limit, the photon-blockade effect observed in the Jaynes-Cummings regime disappears for $ 0.45 \lesssim g \lesssim 1$, which we attributed to a parity shift in the energy spectrum of the Rabi Hamiltonian that induces a cascaded emission of photons. We have also found that for $1\lesssim g/\omega_c \lesssim 2.5$  the second-order autocorrelation function exhibits an oscillatory behavior, leading to a revival of the photon blockade in this range of parameters. This revival, characterized by a strong photon antibunching, takes place when the cascaded decay channel closes while the energy spectrum remains anharmonic. For a even larger coupling strength, the anharmonicity of the spectrum becomes smaller than the dissipation rates, and as a consequence, we observed that the emitted light is coherent. For the dissipation rates considered here, this occurs for $g/\omega_c\gtrsim 2.5$. The regime of parameters considered here is reachable with current technologies and the different regimes of photon emission presented in this paper could therefore be observed in circuit QED experiments.\\ 

This work was supported by the EU Integrating Project SIQS, the EU STREPs DIADEMS and EQUAM, the ERC Synergy Grant BioQ and Alexander von Humboldt Professorship as well as the DFG via the SFB TRR/21 and SPP 1601.



\begin{thebibliography}{99}
\bibitem{Haroche:2006}
S. Haroche and J. M. Raimond, \textit{Exploring the Quantum: Atoms, Cavities and Photons} (Oxford Graduate Texts, New-York, 2006)
\bibitem{Gleyzes:2007}
S. Gleyzes, S. Kuhr, C. Guerlin, J. Bernu, S. Del\'eglise, U. Busk Hoff, Michel Brune, J.-M. Raimond, and S. Haroche, Nature \textbf{446}, 297 (2007)
\bibitem{Imamoglu:1999}
A. Imamoglu, D. D. Awschalom, G. Burkard, D. P. DiVincenzo, D. Loss, M. Sherwin, and A. Small, Phys. Rev. Lett. \textbf{83}, 4204 (1999)
\bibitem{Majer:2007}
J. Majer \textit{et al.}, Nature \textbf{449}, 443 (2007) 
\bibitem{Rempe:1987}
G. Rempe, H. Walther, and N. Klein, Phys. Rev. Lett. \textbf{58}, 353 (1987)
\bibitem{Reithmaier:2004}
J. P. Reithmaier, G. Sek, A. L\"offler, C. Hofmann, S. Kuhn, S. Reitzenstein, L. V. Keldysh, V. D. Kulakovskii, T. L. Reinecke and A. Forchel, Nature \textbf{432}, 197 (2004)
\bibitem{Wallraff:2004}
A. Wallraff, D. I. Schuster, A. Blais, L. Frunzio, R.- S. Huang, J. Majer, S. Kumar, S. M. Girvin and R. J. Schoelkopf, Nature \textbf{431}, 162 (2004)
\bibitem{Peter:2005}
E. Peter, P. Senellart, D. Martrou, A. Lema\^itre, J. Hours, J. M. G\'erard, and J. Bloch, Phys. Rev. Lett. \textbf{95}, 067401 (2005)
\bibitem{Imamoglu:1997}
A. Imamoglu, H. Schmidt, G. Woods, and M. Deutsch, Phys. Rev. Lett. \textbf{79}, 1467 
\bibitem{Birnbaum:2005}
K. M. Birnbaum, A. Boca, R. Miller, A. D. Boozer, T. E. Northup and H. J. Kimble, Nature \textbf{436}, 87 (2005)
\bibitem{Carmichael:2015}
H. J. Carmichael, Phys. Rev. X, \textbf{5}, 031028 (2015)
\bibitem{Verger:2006}
A. Verger, C. Ciuti, and I. Carusotto, Phys. Rev. B \textbf{73}, 193306 (2006)
\bibitem{Liew:2010}
T. C. H. Liew and V. Savona, Phys. Rev. Lett. \textbf{104}, 183601(2010)
\bibitem{Bozyigit:2010}
 D. Bozyigit, C. Lang, L. Steffen, J. M. Fink, C. Eichler, M. Baur,	R. Bianchetti, P. J. Leek, S. Filipp, M. P. da Silva, A. Blais	and A. Wallraff, Nat. Phys. \textbf{7}, 154 (2011)
\bibitem{Lang:2011}
C. Lang, D. Bozyigit, C. Eichler, L. Steffen, J. M. Fink, A. A. Abdumalikov, Jr., M. Baur, S. Filipp, M. P. da Silva, A. Blais, and A. Wallraff, Phys. Rev. Lett. \textbf{106}, 243601(2011)
\bibitem{Hoffman:2011}
A. J. Hoffman, S. J. Srinivasan, S. Schmidt, L. Spietz, J. Aumentado, H. E. T\"ureci, and A. A. Houck, Phys. Rev. Lett. \textbf{107}, 053602 (2011)
\bibitem{Fink:2008}
J. M. Fink, M. G\"oppl, M. Baur, R. Bianchetti, P. J. Leek, A. Blais and A. Wallraff, Nature \textbf{454}, 315 (2008)
\bibitem{Devoret:2007}
M. H. Devoret, S. Girvin and R. Schoelkopf, Ann. Phys.  \textbf{16}, 767 (2007)
\bibitem{Bourassa:2009}
J. Bourassa, J. M. Gambetta, A. A. Abdumalikov, Jr., O. Astafiev, Y. Nakamura, and A. Blais, Phys. Rev. A \textbf{80}, 032109 (2009)
\bibitem{Todorov:2010}
Y. Todorov, A. M. Andrews, R. Colombelli, S. De Liberato, C. Ciuti, P. Klang, G. Strasser, and C. Sirtori, Phys. Rev. Lett. \textbf{105}, 196402 (2010)
\bibitem{Niemczyk:2010}
T. Niemczyk \textit{et al.,} Nat. Phys. \textbf{6}, 772 (2010)
\bibitem{Forn-Diaz:2010}
P. Forn-D\'iaz, J. Lisenfeld, D. Marcos, J. J. GarcÃ­a-Ripoll, E. Solano, C. J. P. M. Harmans, and J. E. Mooij, Phys. Rev. Lett. \textbf{105}, 237001 (2010)
\bibitem{Nataf:2011}
P. Nataf and C. Ciuti, Phys. Rev. Lett. \textbf{107}, 190402 (2011)
\bibitem{Forn-Diaz:2016}
P. Forn-D\'iaz, G. Romero, C. J. P. M. Harmans, E. Solano and J. E. Mooij, Sci. Rep \textbf{6}, 26720 (2016)
\bibitem{Yoshihara:2016}
F. Yoshihara, T. Fuse, S. Ashhab, K. Kakuyanagi, S. Saito and K. Semba, arXiv:1602.00415 (2016)
\bibitem{Irish:2007}
E. K. Irish, Phys. Rev. Lett. \textbf{99}, 173601 (2007)
\bibitem{Ashhab:2010}
S. Ashhab and F. Nori, Phys. Rev. A \textbf{81}, 042311(2010)
\bibitem{Hwang:2010}
M. -J. Hwang and M. -S. Choi, Phys. Rev. A \textbf{82}, 025802 (2010)
\bibitem{Braak:2011}
D. Braak, Phys. Rev. Lett. \textbf{107}, 100401 (2011)
\bibitem{Hwang:2015}
M.-J. Hwang, R. Puebla and M. B. Plenio, Phys. Rev. Lett. \textbf{115}, 180404 (2015)
\bibitem{Ciuti:2006}
C. Ciuti and I. Carusotto, Phys. Rev. A \textbf{74}, 033811(2006)
\bibitem{DeLiberato:2009}
S. De Liberato, D. Gerace, I. Carusotto, and C. Ciuti Phys. Rev. A \textbf{80}, 053810 (2009)
\bibitem{Beaudoin:2011}
F. Beaudoin, J. M. Gambetta and A. Blais, Phys. Rev. A \textbf{84}, 043832 (2011)
\bibitem{Ridolfo:2012}
A. Ridolfo, M. Leib, S. Savasta, and M. J. Hartmann, Phys. Rev. Lett. \textbf{109}, 193602 (2012)
\bibitem{Henriet:2014}
L. Henriet, Z. Ristivojevic, P. P. Orth, and K. Le Hur, Phys. Rev. A \textbf{90}, 023820 (2014)
\bibitem{Larson:2007}
J. Larson, Physica Scripta, \textbf{76}, 146 (2007)
\bibitem{Casanova:2010}
J. Casanova, G. Romero, I. Lizuain, J. J. GarcÃ­a-Ripoll, and E. Solano, Phys. Rev. Lett. \textbf{105}, 263603 (2010)
\bibitem{Hwang:2016}
M. -J. Hwang, M. S. Kim, and M. -S. Choi, Phys. Rev. Lett. \textbf{116}, 153601(2016)
\bibitem{Shamailov:2010}
S.S. Shamailov, A.S. Parkins, M.J. Collett and H.J. Carmichael, Optics Commun. \textbf{283}, 766 (2010)
\bibitem{LeBoite:2014}
A. Le Boit\'e, G. Orso and C. Ciuti, Phys. Rev. A \textbf{90}, 063821 (2014)
\end{thebibliography}
\end{document}